\documentclass[12pt]{article}
\usepackage{amssymb}

\usepackage{graphicx}
\usepackage{layout,amsmath}

\begin{document}

\begin{center}
\textbf{SOLUTION OF MASSLESS SPIN ONE WAVE EQUATION IN
ROBERTSON-WALKER SPACE-TIME}
\end{center}

\bigskip

\bigskip

\begin{center}
YUSUF SUCU \footnote{E-mail : ysucu@sci.akdeniz.edu.tr} and  NURI
UNAL\footnote{E-mail : unal@sci.akdeniz.edu.tr}

\textit{Akdeniz University, Department of Physics, PK 510, 07058
Antalya, Turkey}

\bigskip
\end{center}

\bigskip

\begin{abstract}
We generalize the quantum spinor wave equation for photon into the curved
space-time and discuss the solutions of this equation in Robertson-Walker
space-time and compare them with the solution of the Maxwell equations in
the same space-time.
\end{abstract}

\bigskip

PACS numbers: 03.65.Pm, 98. 80. Cq

\section{\textbf{\ Introduction}}

In 1957 Brill and Wheeler investigated the solution of the Dirac
equation in Schwarzschild metric and they also discussed the
geodesic equation for the photon in the same metric \cite{1}.
Later there are a lot of studies on the solution of the
Klein-Gordon equation in expanding universes. To understand the
contributions of the spin to particle creation and the vacuum
structure of the curved space-times some authors discussed the
solution of the Dirac equation in Robertson-Walker (RW) metrics
\cite{2}-\cite{5}.

Generally the photons are formulated as the quantum of the Maxwell
fields in flat and curved space-times. There are continuous
attempts to write the Maxwell equations as the spinor equations
and also to represent the photon by a quantum wave equation for
photon \cite{6}-\cite{13}. All massless spinning particle wave
equations can be written as the spinor equations in the
Newmann-Penrose formalism\cite{14}.

The spin is introduced in Dirac equation and there are a lot of
models to discuss the \ the classical analogue of this new degree
of freedom\cite{15}. In 1984, Barut and Zanghi proposed a
classical model of the zitterbewegung and discussed the free
particle trajectories in this model\cite{16}. The quantization of
this model is discussed from different aspects\cite{17} and it
gives the Dirac equation for the spin-1/2 electron, the
Duffin-Kemer-Petiau equation for the spin-1 particle as well as
the higher-spinwave equations\cite{18}.

Recently one of us proposed a classical model of the
zitterbewegung for the massless spinning particles\cite{19} and
the quantization of this model gives the spinor wave equations for
the massless spinning particle as well as two component neutrino
equation.

The aim of this paper is to discuss the generalization of the classical
zitterbewegung model, to obtain the quantum wave equations for photons in
curved background metric, and as an example, to discuss the solutions of
this wave equation in the RW metrics.

In section two we discuss the generalization of the simple model
of the zitterbewegung to the curved space-times and the
quantization of this classical system and derive the Weyl equation
and spin-1 analogue of it in the curved space-times. In section
four \ we discuss the solution of this spin-1 wave equation for
the Robertson- Walker space-times. For the comparison we also
obtain the solutions of the Maxwell equations for the same metrics
in Appendix.

\section{\textbf{Classical and quantum system}}

The classical model of the zitterbewegung for electron is formulated in the
configuration space $M^{4}\otimes C^{4}$ with the canonically conjugate
internal and external coordinates and momenta of the particle:$\left(
x,p\right) $ and $\left( \eta ^{\dagger },\eta \right) .$ Here $\eta
^{\dagger }$and $\eta $ are the $4$ component complex spinors.For the
massless particles the action of the classical spinning particle in curved
space-time is
\begin{equation}
A=\int ds\left[ \frac{1}{2i}\left( \frac{d\eta ^{\dagger }}{ds}\eta -\eta
^{\dagger }\frac{d\eta }{ds}\right) +p_{\mu }\frac{dx^{\mu }}{ds}-\eta
^{\dagger }\sigma ^{\mu }\eta \left( p_{\mu }+i\eta ^{\dagger }\Gamma _{\mu
}\eta \right) \right] ,
\end{equation}%
where $\sigma ^{\alpha }$ is $2\times 2$ Pauli matrices, $\eta (s)$ is a two
component complex spinor, $\Gamma _{\mu }$ is the spin connection and given
in next section $p^{\mu }$ and $\ x^{\mu }$ are the canonical conjugate,
external coordinates of the particle. The internal dynamics of the particle
is described by the $\eta ^{\dagger }$ and $\eta $, which are the canonical
conjugate, holomorphic coordinates. In the second term of the Lagrangian the
internal and external dynamics are coupled by the Lagrange multiplier $p.$
The internal coordinates describe the four coupled oscillators. We choose $%
\hslash $\ $=c=1$.The classical equations of motion for $x^{\mu },$ $p^{\mu
},$ $\eta ^{\dagger }$ and $\eta $ are derived in Reference [19].

The elements of the configuration space of the spinning particle are $x^{\mu
}$ and $\eta ^{\dagger }.$ Then the Schr\"{o}dinger wave function is $\Phi
\left( x,\eta ^{\dagger };s\right) $ and it satisfies the following wave
equation:
\begin{equation}
i\frac{\partial }{\partial s}\Phi \left( x,\eta ^{\dagger };s\right) =%
\widehat{H}\Phi \left( x,\eta ^{\dagger };s\right) ,
\end{equation}%
where $\widehat{H}$ is the Hamiltonian and given by
\begin{equation}
\widehat{H}=\widehat{\eta ^{\dagger }}\gamma ^{\mu }\widehat{\eta }\left(
\widehat{p}_{\mu }+i\widehat{\eta ^{\dagger }}\Gamma _{\mu }\widehat{\eta }%
\right) .
\end{equation}%
$\widehat{H}$ is the function of the canonical conjugate variables $\widehat{%
\eta ^{\dagger }}$ and $\widehat{\eta }$ and $\widehat{p}^{\mu }$ and $%
\widehat{x}^{\mu }.$ $\ \widehat{\eta }$ and $\widehat{p}^{\mu }$ are
represented as
\begin{equation*}
\widehat{\eta }=\frac{\partial }{\partial \eta ^{\dagger }},
\end{equation*}%
\begin{equation}
\widehat{p}^{\mu }=i\frac{\partial }{\partial x_{\mu }}.
\end{equation}%
$\Phi \left( x,\eta ^{\dagger };s\right) $ represents all eigenvalues of the
spin. To obtain the wave equations for certain eigenvalues of the spin we
expand $\Phi $ as a power series of $\eta ^{\dagger }:$%
\begin{equation}
\Phi \left( x,\eta ^{\dagger };s\right) =\Psi _{0}\left( x,s\right) +\eta
_{\alpha }^{\dagger }\Psi _{\alpha }\left( x,s\right) +\frac{1}{2}\left(
\eta _{\alpha }^{\dagger }\eta _{\beta }^{\dagger }+\eta _{\beta }^{\dagger
}\eta _{\alpha }^{\dagger }\right) \Psi _{\alpha \beta }\left( x,s\right)
+...
\end{equation}%
In this expansion $\Psi _{\alpha \beta \gamma ...}$ is symmetric
under the exchange of the indices and thus for example $\Psi
_{\alpha \beta }$ has $3$ independent components. We substitute
this expansion into the Schr\"{o}dinger equation (2). Then each
power of $\eta ^{\dagger }$ satisfies the separate wave equations:
\begin{equation*}
\left[ i\frac{\partial }{\partial s}-\frac{1}{2\Lambda }p_{\mu }g^{\mu
\upsilon }p_{\upsilon }\right] \Psi _{\circ }\left( x,s\right) =0,
\end{equation*}%
\begin{equation}
\left[ i\frac{\partial }{\partial s}-\sigma ^{\mu }\left( \widehat{p}_{\mu
}-i\Gamma _{\mu }\right) \right] _{\alpha \beta }\Psi _{\alpha \beta }\left(
x,s\right) =0,
\end{equation}%
\begin{equation*}
\{i\frac{\partial }{\partial s}I\otimes I-\Sigma ^{\mu }\left( x\right) %
\left[ p_{\mu }+i\Gamma _{\mu }\left( x\right) \otimes \text{I}+\text{I}%
\otimes i\Gamma _{\mu }\left( x\right) \right] \}_{\alpha _{1}\alpha
_{2},\beta _{1}\beta _{2}}\Psi _{\beta _{1}\beta _{2}}=0.
\end{equation*}%
The first, second and third of these equations are the Klein
Gordon equation with a parameter $\Lambda $, the neutrino equation
and the photon wave equations in general coordinate frame
respectively and the $\Sigma ^{\mu }\left( x\right) $ in the third
equation is

\begin{equation}
\Sigma ^{\mu }\left( x\right) =\sigma ^{\mu }\left( x\right) \otimes \text{I}%
+\text{I}\otimes \sigma ^{\mu }\left( x\right) .
\end{equation}%
\noindent In flat Minkowski space-time the photon wave equation becomes
\begin{equation}
\{i\frac{\partial }{\partial s}I\otimes I-\widehat{p}_{\mu }\left( \sigma
^{\mu }\otimes I+I\otimes \sigma ^{\mu }\right) \}_{\alpha _{1}\alpha
_{2},\beta _{1}\beta _{2}}\Psi _{\beta _{1}\beta _{2}}=0.
\end{equation}%
\noindent For the massless particles $\frac{\partial }{\partial
s}=0$ and Eq.(7) becomes
\begin{equation}
2iI\otimes I\frac{\partial \Psi }{\partial t}=\widehat{\overset{%
\rightharpoonup }{p}}\left[ \overset{\rightharpoonup }{\sigma }\otimes
I+I\otimes \overset{\rightharpoonup }{\sigma }\right] \Psi =0.
\end{equation}%
\noindent In equation (9) $\Psi $ has three components. We write
this
equation explicitly by introducing a vector wave function $\overset{%
\rightharpoonup }{\Psi }$ such that
\begin{equation}
\overset{\rightharpoonup }{\Psi }=\left( \Psi _{-},\sqrt{2}\Psi _{0},\Psi
_{+}\right) ,
\end{equation}%
\noindent where $\Psi _{-,+,0}$ are the components of the spinor
$\Psi .$ Then Eq.(9) becomes
\begin{equation*}
i\frac{\partial \overset{\rightharpoonup }{\Psi }}{\partial t}=\nabla \times
\overset{\rightharpoonup }{\Psi .}
\end{equation*}%
\noindent This is the six of the Maxwell equations and if $\overset{%
\rightharpoonup }{\Psi }$ is time dependent, then also $\nabla \overset{%
\rightharpoonup }{\Psi }=0.$ Thus the spinor equation (6) (for
photon) which is obtained from the generalization of the Eq.(9) to
curved space-time is equivalent to the generalization of Maxwell
equations to curved space-time. Since this equation has only first
order derivatives with respect to space and times we define $\Psi
^{\dagger }\sigma ^{\mu }\Psi $ as the probability current for the
photon.

\section{\textbf{Solution of the photon wave equation in RW metrics}}

The RW metric is
\begin{equation}
ds^{2}=g_{\mu \nu }dx^{\mu }dx^{\nu }=dt^{2}-a^{2}\left( t\right) \left(
\frac{dr^{2}}{1-kr^{2}}+r^{2}d\theta ^{2}+r^{2}\sin ^{2}\theta d\varphi
^{2}\right) ,
\end{equation}%
where \ $a^{2}\left( t\right) $ is the expansion parameter and $k$
is the curvature parameter between $-1$ and $1.$ The metric tensor
can be represented in terms of the symmetric form of the vierbein
$L_{\mu }^{\alpha
}$%
\begin{equation}
g_{\mu \nu }=L_{\mu }^{\alpha }L_{\nu }^{\beta }\eta _{\alpha \beta },
\end{equation}%
where $\eta _{\alpha \beta }$ is metric of the Minkowski frame and $\eta
_{\alpha \beta }=\left( +1,-1,-1,-1\right) .$ The vierbein of the metric is
\begin{equation*}
L_{\circ }^{\circ }=1,\text{ \ \ }L_{\circ }^{i}=0
\end{equation*}%
and
\begin{equation}
L_{j}^{i}=L_{i}^{j}=\frac{a}{r}\delta _{ij}+a\frac{\left( 1-\rho \right) }{%
\rho r^{2}}x^{i}x^{j},
\end{equation}%
where $\rho $ is $\left( 1-kr^{2}\right) ^{%
{\frac12}%
}$. Then the Dirac matrices are related to each other by
\begin{equation*}
\sigma ^{\mu }\left( x\right) =L_{\alpha }^{\mu }\left( x\right) \sigma
^{\alpha }=g^{\mu \nu }L_{\nu }^{\alpha }\eta _{\alpha \beta }\sigma ^{\beta
}
\end{equation*}%
and
\begin{equation}
\sigma _{\mu }\left( x\right) =L_{\mu }^{\alpha }\left( x\right) \sigma
_{\alpha },
\end{equation}%
where $\sigma ^{\mu }\left( x\right) $ and $\sigma ^{\alpha }$ are Pauli
matrices in the general coordinate frame and the Minkowski frame
respectively and they satisfy the following anticommutation relations:
\begin{equation*}
\sigma ^{\alpha }\sigma ^{\beta }+\sigma ^{\beta }\sigma ^{\alpha }=2\eta
^{\alpha \beta },
\end{equation*}%
\begin{equation}
\sigma ^{\alpha }\left( x\right) \sigma ^{\beta }\left( x\right) +\sigma
^{\beta }\left( x\right) \sigma ^{\alpha }\left( x\right) =2g^{\alpha \beta
}\left( x\right) .
\end{equation}%
By using the expression of \ $L_{\alpha }^{\mu }\left( x\right) $ \ we write
$\sigma ^{\mu }\left( x\right) $ \ as
\begin{equation*}
\sigma ^{\circ }\left( x\right) =\sigma ^{\circ },
\end{equation*}%
\begin{equation}
\overrightarrow{\sigma }\left( x\right) =-\frac{1}{a}\left[ \overrightarrow{%
\sigma }-\frac{1-\rho }{r^{2}}\overrightarrow{r}\overrightarrow{r}\cdot
\overrightarrow{\sigma }\right] .
\end{equation}%
The spin connection is derived by the following expression:
\begin{equation*}
\Gamma _{\mu }\left( x\right) =-\frac{1}{8}\left[ \sigma ^{\nu }\left(
x\right) ,\sigma _{\nu ;\mu }\left( x\right) \right] .
\end{equation*}%
It is
\begin{equation*}
\Gamma ^{\circ }=0
\end{equation*}%
\begin{equation}
\overrightarrow{\Gamma }=-\frac{1}{2}\left[ \dot{a}\left( \sigma ^{\circ }%
\overrightarrow{\sigma }+\frac{1-\rho }{\rho r^{2}}\overrightarrow{r}\sigma
^{\circ }\overrightarrow{r}\cdot \overrightarrow{\sigma }\right) -i\frac{%
\left( 1-\rho \right) }{r^{2}}\left( \overrightarrow{\sigma }\times
\overrightarrow{r}\right) \right] .
\end{equation}%
\noindent Then the interaction term becomes
\begin{equation*}
-[\overrightarrow{\sigma }\left( x\right) \otimes \text{I}+\text{I}\otimes
\overrightarrow{\sigma }\left( x\right) ]\cdot \left[ \overrightarrow{\Gamma
}\left( x\right) \otimes \text{I}+\text{I}\otimes \overrightarrow{\Gamma }%
\left( x\right) \right] =
\end{equation*}%
\begin{equation*}
-\frac{1}{2a}\{\left[ \overrightarrow{\sigma }-\left( 1-\rho \right)
\overrightarrow{r}\overrightarrow{\sigma }\cdot \overrightarrow{r}\right]
\otimes \text{I}+\text{I}\otimes \left[ \overrightarrow{\sigma }-\left(
1-\rho \right) \overrightarrow{r}\overrightarrow{\sigma }\cdot
\overrightarrow{r}\right] \}
\end{equation*}%
\begin{equation*}
\times \{\left[ \dot{a}\left( \overrightarrow{\sigma }+\frac{1-\rho }{\rho
r^{2}}\overrightarrow{r}\overrightarrow{\sigma }\cdot \overrightarrow{r}%
\right) -i\frac{1-\rho }{r^{2}}\left( \overrightarrow{\sigma }\times
\overrightarrow{r}\right) \right] \otimes \text{I}
\end{equation*}%
\begin{equation}
+\text{I}\otimes \left[ \dot{a}\left( \overrightarrow{\sigma }+\frac{1-\rho
}{\rho r^{2}}\overrightarrow{r}\overrightarrow{\sigma }\cdot \overrightarrow{%
r}\right) -i\frac{1-\rho }{r^{2}}\left( \overrightarrow{\sigma }\times
\overrightarrow{r}\right) \right] \}.
\end{equation}%
We will integrate the time and the angular coordinates of the wave function
by using the group theoretical methods\cite{20}. For this reason we derive
the photon wave equation from the variation of the following action with
respect to $\Phi ^{\dagger }\left( \overrightarrow{r},t\right) $:
\begin{equation*}
A=\int d^{4}x\sqrt{-g}\Phi ^{\dagger }\left( \overrightarrow{x},t\right) \{2i%
\text{I}\otimes \text{I}\frac{\partial }{\partial t}
\end{equation*}%
\begin{equation}
-\overrightarrow{\Sigma }\left( x\right) \left[ \overrightarrow{p}+i%
\overrightarrow{\Gamma }\left( x\right) \otimes \text{I}+\text{I}\otimes i%
\overrightarrow{\Gamma }\left( x\right) \right] \}\Phi \left(
\overrightarrow{x},t\right) .
\end{equation}%
Because of the spherical symmetry of the intersection terms we
write the action in spherical coordinates $\left( r,\theta
,\varphi \right) $. It is
\begin{equation*}
A=\int dtdrd\theta d\varphi \sqrt{-g}\Phi ^{\dagger }[2i\text{I}\otimes
\text{I}\frac{\partial }{\partial t}+i\frac{\dot{a}}{a}\left( \Sigma
^{r}\Sigma ^{r}+\Sigma ^{\theta }\Sigma ^{\theta }+\Sigma ^{\varphi }\Sigma
^{\varphi }\right)
\end{equation*}%
\begin{equation}
+\frac{i}{a}\left( \rho \Sigma ^{r}\frac{\partial }{\partial r}+\frac{1}{r}%
\Sigma ^{\theta }\frac{\partial }{\partial \theta }+\frac{1}{r\sin \theta }%
\Sigma ^{\varphi }\frac{\partial }{\partial \varphi }\right) -i\frac{\left(
1-\rho \right) }{a}\frac{\Sigma ^{r}}{r}]\Phi ,
\end{equation}%
where $\overset{.}{a}=\frac{da}{dt},$ $\sqrt{-g}$ is
\begin{equation}
\sqrt{-g}=\frac{a^{3}}{1-kr^{2}}r^{2}\sin \theta ,
\end{equation}%
and $\Sigma ^{r},$ $\Sigma ^{\theta }$ \ and $\Sigma ^{\varphi }$ are the
components of the $\overrightarrow{\Sigma }$ along the axis $r,$ $\theta $
and $\varphi .$ We rewrite the action in equation(20) as
\begin{equation*}
A=i\int d\tau d\chi d\theta d\varphi e^{\frac{3\alpha }{2}}\xi ^{2}\left(
\chi \right) \sin \theta \Phi ^{\dagger }\left( \chi ,\theta ,\varphi ;\tau
\right) [2\text{I}\otimes \text{I}\left( \frac{\partial }{\partial \tau }+%
\frac{d\alpha }{d\tau }\right) +
\end{equation*}%
\begin{equation}
\Sigma ^{\chi }\left( \frac{\partial }{\partial \chi }+\frac{\xi ^{\prime
}\left( \chi \right) -1}{\xi \left( \chi \right) }\right) +\frac{1}{\xi
\left( \chi \right) }\left( \Sigma ^{\theta }\frac{\partial }{\partial
\theta }+\frac{1}{\sin \theta }\Sigma ^{\varphi }\frac{\partial }{\partial
\varphi }\right) ]\Phi \left( \chi ,\theta ,\varphi ;\tau \right) ,
\end{equation}%
where the new coordinates $\tau $ and $\chi $ are defined as
\begin{equation}
d\tau =e^{\frac{\alpha \left( \tau \right) }{2}}dt=a\left( t\right) dt,\
r=\xi \left( \chi \right) =\frac{1}{\sqrt{k}}\sin \sqrt{k}\chi \ \ \ \text{%
and \ }\ \xi ^{\prime }\left( \chi \right) =\frac{d\xi }{d\chi }.
\end{equation}%
Then the metric becomes
\begin{equation}
ds^{2}=e^{\alpha \left( \tau \right) }\left[ d\tau ^{2}-d\chi ^{2}-\xi
^{2}\left( \chi \right) \left( d\theta ^{2}+\sin ^{2}\theta \text{ }d\varphi
^{2}\right) \right] .
\end{equation}%
To eliminate the $\tau $ dependent potentials we introduce a new function $f$
\ such that
\begin{equation}
\Phi \left( \chi ,\theta ,\varphi ,\tau \right) =\xi ^{-1}\left( \chi
\right) \ e^{-\alpha }f\left( \chi ,\theta ,\varphi ;\tau \right) .
\end{equation}%
Then the action becomes
\begin{equation*}
A=i\int d\tau e^{-\frac{\alpha }{2}}d\chi d\theta d\varphi \sin \theta
f^{\dagger }\left( \chi ,\theta ,\varphi ;\tau \right) [2(\text{I}\otimes
\text{I})\frac{\partial }{\partial \tau }+
\end{equation*}%
\begin{equation}
\Sigma ^{\chi }\left( \frac{\partial }{\partial \chi }-\frac{1}{\xi \left(
\chi \right) }\right) +\frac{1}{\xi \left( \chi \right) }\left( \Sigma
^{\theta }\frac{\partial }{\partial \theta }+\frac{1}{\sin \theta }\Sigma
^{\varphi }\frac{\partial }{\partial \varphi }\right) ]f\left( \chi ,\theta
,\varphi ;\tau \right) .
\end{equation}%
We define a rotation from the $\widehat{r}$ to $x^{3}-$axis. Under this
rotation the spinors $f\left( \chi ,\theta ,\varphi ;\tau \right) $ rotate
as
\begin{equation}
f\left( \chi ,\theta ,\varphi ;\tau \right) \rightarrow e^{i\omega \tau
}SF\left( \chi ,\theta ,\varphi \right) ,
\end{equation}%
where $S$ is the rotation operator of the spinors and defined as
\begin{equation}
S=e^{-\frac{i}{2}\Sigma ^{2}\theta -\frac{i}{2}\Sigma ^{1}\varphi .}
\end{equation}%
Then the action becomes
\begin{equation*}
A=i\int d\tau e^{-\frac{\alpha }{2}}d\chi d\theta d\varphi \sin \theta
F^{\dagger }\left( \chi ,\theta ,\varphi \right) \{2i\omega +\Sigma
^{3}\left( \frac{\partial }{\partial \chi }-\frac{1}{\xi \left( \chi \right)
}\right) +
\end{equation*}%
\begin{equation}
\frac{1}{\xi \left( \chi \right) }[\frac{1}{2}\left( \Sigma ^{1}+i\Sigma
^{2}\right) \partial _{+}+\frac{1}{2}\left( \Sigma ^{1}-i\Sigma ^{2}\right)
\partial _{-}-\frac{i}{2}\left[ \Sigma ^{1},\Sigma ^{2}\right] ]\}F\left(
\chi ,\theta ,\varphi \right) ,
\end{equation}%
where we have used the following properties of the $S$ operators:
\begin{equation*}
S^{-1}\frac{\partial }{\partial \tau }S=\frac{\partial }{\partial \tau },
\end{equation*}%
\begin{equation*}
S^{-1}\Sigma ^{\chi }\frac{\partial }{\partial \chi }S=\Sigma ^{3}\frac{%
\partial }{\partial \chi },
\end{equation*}%
\begin{equation}
S^{-1}\Sigma ^{\theta }\frac{\partial }{\partial \theta }S=\Sigma ^{1}\left(
\frac{\partial }{\partial \theta }-i\Sigma ^{2}\right) ,
\end{equation}%
\begin{equation*}
S^{-1}\Sigma ^{\varphi }\frac{1}{\sin \varphi }\frac{\partial }{\partial
\varphi }S=\Sigma ^{2}\left( \frac{1}{\sin \varphi }\frac{\partial }{%
\partial \varphi }+\frac{i}{2}\Sigma ^{1}-\frac{i}{2}\Sigma ^{3}\cot \theta
\right) ,
\end{equation*}%
and $\partial _{\pm }$ are
\begin{equation}
\partial _{\pm }=\mp \frac{\partial }{\partial \theta }+\frac{i}{\sin \theta
}\frac{\partial }{\partial \varphi }+\frac{1}{2}\Sigma ^{3}\cot \theta .
\end{equation}%
The $\partial _{\pm }$ are the rising and lowering operators of
the angular momentum eigenfunctions
\begin{equation}
D_{\lambda ,m}^{j}\left( \theta ,\varphi \right) =\left\langle \lambda
\left| R\left( \theta ,\varphi \right) \right| jm\right\rangle .
\end{equation}%
We expand the spinors $F$ in terms of the angular momentum eigenfunctions as
\begin{equation}
F\left( \chi ,\theta ,\varphi \right) =4\pi \underset{jm}{\Sigma }\left(
2j+1\right) \left(
\begin{array}{c}
F_{+}^{jm}\left( \chi \right) D_{+1,m}^{j}\left( \theta ,\varphi \right)  \\
\begin{array}{c}
F_{\circ }^{jm}\left( \chi \right) D_{\circ ,m}^{j}\left( \theta ,\varphi
\right)  \\
F_{\circ }^{jm}\left( \chi \right) D_{\circ ,m}^{j}\left( \theta ,\varphi
\right)
\end{array}
\\
F_{-}^{jm}\left( \chi \right) D_{-1,m}^{j}\left( \theta ,\varphi \right)
\end{array}%
\right) .
\end{equation}%
\noindent Then we calculate the operation of \ $\partial _{\pm }$ on \ $%
F\left( \chi ,\theta ,\varphi \right) $ \ by using the following rules:
\begin{equation}
\partial _{\pm }D_{\lambda ,m}^{j}\left( \theta ,\varphi \right) =\left[
\left( j\pm \lambda +1\right) \left( j\pm \lambda \right) \right] ^{%
{\frac12}%
}D_{\lambda \pm 1,m}^{j}\left( \theta ,\varphi \right) .
\end{equation}%
We perform the matrix operations and the angular integrations by using the
orthogonality relations of the $\ D_{\lambda ,m}^{j}\left( \theta ,\varphi
\right) $ functions. The result is the following 2-dimensional form of the
action:
\begin{equation*}
A=2i\left( 4\pi \right) ^{3}\underset{jm}{\Sigma }\left( 2j+1\right) \int
d\tau e^{-\frac{\alpha }{2}}d\chi F_{jm}^{\dagger }\left( \chi \right)
[2i\omega +
\end{equation*}%
\begin{equation}
\Sigma ^{3}\frac{\partial }{\partial \chi }-\frac{i}{\xi \left( \chi \right)
}\sqrt{j\left( j+1\right) }\Sigma ^{2}]F_{jm}\left( \chi \right) ,
\end{equation}%
where $F_{jm}\left( \chi \right) $ is
\begin{equation}
F_{jm}\left( \chi \right) =\left(
\begin{array}{c}
F_{+}^{jm}\left( \chi \right)  \\
\begin{array}{c}
F_{\circ }^{jm}\left( \chi \right)  \\
F_{\circ }^{jm}\left( \chi \right)  \\
F_{-}^{jm}\left( \chi \right)
\end{array}%
\end{array}%
\right) .
\end{equation}%
The variation of the action with respect to $F_{+}^{jm}\left( \chi \right) $
gives the following radial equations for the photon:
\begin{equation}
\lbrack i\omega +\frac{1}{2}\Sigma ^{3}\frac{\partial }{\partial \chi }-%
\frac{i}{2\xi \left( \chi \right) }\sqrt{j\left( j+1\right) }\Sigma
^{2}]F_{jm}\left( \chi \right) =0.
\end{equation}%
By using the explicit form of the Pauli matrices we write this equation for
the components of the spinor $F.$ These are
\begin{equation*}
\left( i\omega +\frac{\partial }{\partial \chi }\right) F_{+}\left( \chi
\right) -\frac{1}{\xi \left( \chi \right) }\sqrt{j\left( j+1\right) }%
F_{\circ }\left( \chi \right) =0,
\end{equation*}%
\begin{equation}
\left( i\omega -\frac{\partial }{\partial \chi }\right) F_{-}\left( \chi
\right) +\frac{1}{\xi \left( \chi \right) }\sqrt{j\left( j+1\right) }%
F_{\circ }\left( \chi \right) =0,
\end{equation}%
\begin{equation*}
i\omega F_{\circ }\left( \chi \right) -\frac{1}{2\xi \left( \chi \right) }%
\sqrt{j\left( j+1\right) }\left[ F_{-}\left( \chi \right) -F_{+}\left( \chi
\right) \right] =0.
\end{equation*}%
Combination of these three equations give
\begin{equation}
\left[ \omega ^{2}+\frac{\partial ^{2}}{\partial \chi ^{2}}-\frac{j\left(
j+1\right) }{\xi ^{2}\left( \chi \right) }\right] \left[ F_{-}\left( \chi
\right) -F_{+}\left( \chi \right) \right] =0.
\end{equation}%
Since $\xi \left( \chi \right) =\frac{1}{\sqrt{k}}\sin \sqrt{k}\chi ,$ the
regular solution of this equation at the origin is given by
\begin{equation}
\left[ F_{-}\left( \chi \right) -F_{+}\left( \chi \right) \right] ={C\left(
\frac{\sin \sqrt{k}\chi }{\sqrt{k}}\right) ^{j+1}}\,{}_{2}F_{1}(\alpha
,\beta ,\gamma ;\sin ^{2}\sqrt{k}\chi ),
\end{equation}%
where $C$ is the normalization constant and $\alpha ,$ $\beta $ and $\gamma $
are $\alpha =\frac{1}{2}\left( j+1+\frac{\omega }{k}\right) $, $\beta =\frac{%
1}{2}\left( j+1-\frac{\omega }{k}\right) $, $\gamma =j+\frac{3}{2}$. Then
the spinor $\Phi $ becomes
\begin{equation*}
\Phi _{\pm }=C\frac{e^{i\omega \tau }}{2a^{2}\left( t\right) }\left( \frac{%
\sin \sqrt{k}\chi }{\sqrt{k}}\right) ^{j+1}\times
\end{equation*}%
\begin{equation*}
\{\left[ 1\pm \frac{i\sqrt{k}}{\omega }\left( j+1\right) \cot \sqrt{k}\chi %
\right] \text{ }_{2}F_{1}\left( \alpha ,\beta ,\gamma ;\sin ^{2}\sqrt{k}\chi
\right) \pm
\end{equation*}%
\begin{equation*}
\frac{i\sqrt{k}}{2\omega }\frac{\alpha \beta }{\gamma }\sin 2\left( \sqrt{k}%
\chi \right) \text{ }_{2}F_{1}\left( \alpha +1,\beta +1,\gamma +1;\sin ^{2}%
\sqrt{k}\chi \right) \},
\end{equation*}%
\begin{equation}
\Phi _{\circ }=iC\frac{e^{i\omega \tau }}{\omega a^{2}\left( t\right) }\sqrt{%
j\left( j+1\right) }\left( \frac{\sin \sqrt{k}\chi }{\sqrt{k}}\right)
^{j}\,{}_{2}F_{1}(\alpha ,\beta ,\gamma ;\sin ^{2}\sqrt{k}\chi ),
\end{equation}%
where the conformal time $\tau $ is related to physical time $t$ by
\begin{equation}
\tau =\int^{t}\frac{dt^{\prime }}{a\left( t^{\prime }\right) }.
\end{equation}

\section{\textbf{Conclusion}}

In this paper we derived the massless particle limit of the Duffin-Kemer
-Petiau equation for the spin-1 particle in the curved space and found the
exact solution of this equation in the RW expanding universes. Because of
the conformal and rotational symmetries of the metric we eliminated the time
and angular coordinates by using the group theoretical methods and reduced
the problem into the radial coordinates.

Since the particle is massless the dependence of the wave
functions on the physical or parametric time do not depend on the
explicit form of the expansion factor $a\left( t\right) .$ Then in
the current density $\Psi ^{\dagger }\sigma \Psi $ there are no
time oscillations, no particle creation for the massless particle
\cite{5}.


\section*{Appendix A}

Here we solve the Maxwell equations in RW metrics for the
completeness. We use the metric in Eq.(24). Then the \ nonzero
component of vierbeins are

\begin{equation}
L_{\tau }^{\circ }=e^{\frac{\alpha }{2}},L_{\chi }^{1}=e^{\frac{\alpha }{2}%
},L_{\theta }^{2}=e^{\frac{\alpha }{2}}\xi ,L_{\varphi }^{3}=e^{\frac{\alpha
}{2}}\xi \sin \theta .  \tag{A1}
\end{equation}%
\noindent Then we can write the contravariant field strengths $F^{\mu \nu }$
in the general coordinate as
\begin{equation*}
F^{01}=e^{-\alpha }E^{1},
\end{equation*}%
\begin{equation*}
F^{02}=\frac{e^{-\alpha }}{\xi }E^{2},
\end{equation*}%
\begin{equation}
F^{03}=\frac{e^{-\alpha }}{\xi \sin \theta }E^{3},  \tag{A2}
\end{equation}%
\begin{equation*}
F^{12}=\frac{e^{-\alpha }}{\xi }B^{3},
\end{equation*}%
\begin{equation*}
F^{23}=\frac{e^{-\alpha }}{\xi ^{2}\sin \theta }B^{1},
\end{equation*}%
\begin{equation*}
F^{31}=\frac{e^{-\alpha }}{\xi \sin \theta }B^{2},
\end{equation*}%
\noindent where $E^{i}$ and $B^{i}$ are the components of the electric and
magnettic fields in the local Lorentz frame. Covariant components of the
field strength tensor are
\begin{equation*}
F_{01}=-e^{\alpha }E^{1},
\end{equation*}%
\begin{equation}
F_{02}=-e^{\alpha }\xi E^{2},  \tag{A3}
\end{equation}%
\begin{equation*}
F_{03}=-e^{\alpha }\xi \sin \theta E^{3},
\end{equation*}%
\noindent and
\begin{equation*}
F_{12}=e^{\alpha }\xi B^{3},
\end{equation*}%
\begin{equation}
F_{23}=e^{\alpha }\xi ^{2}\sin \theta B^{1},  \tag{A4}
\end{equation}%
\begin{equation*}
F_{31}=e^{\alpha }\xi \sin \theta B^{2}.
\end{equation*}%
\noindent The Maxwell equations in the free space are
\begin{equation}
\frac{1}{\sqrt{-g}}\left( \sqrt{-g}F^{\mu \nu }\right) ,_{\nu }=0,  \tag{A5}
\end{equation}%
\noindent and
\begin{equation}
F_{\mu \nu ,\sigma }+F_{\sigma \mu ,\nu }+F_{\nu \sigma ,\mu }=0.  \tag{A6}
\end{equation}%
\noindent In terms of the components these are
\begin{equation*}
\frac{1}{\xi ^{2}}\frac{\partial }{\partial \chi }\left[ \xi ^{2}\left(
\begin{array}{c}
E^{1} \\
B^{1}%
\end{array}%
\right) \right] +\frac{1}{\xi \sin \theta }\frac{\partial }{\partial \theta }%
\left[ \sin \theta \left(
\begin{array}{c}
E^{2} \\
B^{2}%
\end{array}%
\right) \right] +\frac{1}{\xi \sin \theta }\frac{\partial }{\partial \varphi
}\left(
\begin{array}{c}
E^{3} \\
B^{3}%
\end{array}%
\right) =0,
\end{equation*}%
\begin{equation*}
\frac{1}{\xi \sin \theta }\{\frac{\partial }{\partial \theta }\left[ \sin
\theta \left(
\begin{array}{c}
E^{3} \\
B^{3}%
\end{array}%
\right) \right] -\frac{\partial }{\partial \varphi }\left(
\begin{array}{c}
E^{2} \\
B^{2}%
\end{array}%
\right) \}-e^{-\alpha }\frac{\partial }{\partial \tau }\left[ e^{\alpha
}\left(
\begin{array}{c}
-B^{1} \\
E^{1}%
\end{array}%
\right) \right] =0,
\end{equation*}%
\begin{equation}
\frac{1}{\xi }\{\frac{1}{\sin \theta }\frac{\partial }{\partial \varphi }%
\left(
\begin{array}{c}
E^{1} \\
B^{1}%
\end{array}%
\right) -\frac{\partial }{\partial \chi }\left[ \xi \left(
\begin{array}{c}
E^{3} \\
B^{3}%
\end{array}%
\right) \right] \}-e^{-\alpha }\frac{\partial }{\partial \tau }\left[
e^{\alpha }\left(
\begin{array}{c}
-B^{2} \\
E^{2}%
\end{array}%
\right) \right] =0,  \tag{A7}
\end{equation}%
\begin{equation*}
\frac{1}{\xi }\{\frac{\partial }{\partial \chi }\left[ \xi \left(
\begin{array}{c}
E^{2} \\
B^{2}%
\end{array}%
\right) \right] -\frac{\partial }{\partial \theta }\left(
\begin{array}{c}
E^{1} \\
B^{1}%
\end{array}%
\right) \}-e^{-\alpha }\frac{\partial }{\partial \tau }\left[ e^{\alpha
}\left(
\begin{array}{c}
-B^{3} \\
E^{3}%
\end{array}%
\right) \right] =0.
\end{equation*}%
\noindent The complex spinors $\eta $ and $\eta ^{\dagger }$ are defined as
\begin{equation}
\eta =\left(
\begin{array}{c}
E^{1}+iB^{1} \\
E^{3}+iB^{3} \\
E^{2}+iB^{2}%
\end{array}%
\right) ,  \tag{A8}
\end{equation}%
\noindent and
\begin{equation}
\eta ^{\dagger }=\left( E^{1}-iB^{1}\text{ , }E^{3}-iB^{3}\text{ , }%
E^{2}-iB^{2}\right) .  \tag{A9}
\end{equation}%
\noindent Then the spinor form of the Maxwell equations are
\begin{equation*}
\frac{1}{\xi ^{2}}\frac{\partial }{\partial \chi }\left( \xi ^{2}\eta
^{1}\right) +\frac{1}{\xi \sin \theta }\frac{\partial }{\partial \theta }%
\left( \sin \theta \text{ }\eta ^{2}\right) +\frac{1}{\xi \sin \theta }\frac{%
\partial \eta ^{3}}{\partial \varphi }=0,
\end{equation*}%
\begin{equation*}
\frac{1}{\xi }\left[ \frac{1}{\sin \theta }\frac{\partial }{\partial \theta }%
\left( \sin \theta \text{ }\eta ^{3}\right) -\frac{1}{\sin \theta }\frac{%
\partial \eta ^{2}}{\partial \varphi }\right] -ie^{-\alpha }\frac{\partial }{%
\partial \tau }\left( e^{\alpha }\eta ^{1}\right) =0,
\end{equation*}%
\begin{equation}
\frac{1}{\xi }\left[ \frac{1}{\sin \theta }\frac{\partial \eta ^{1}}{%
\partial \varphi }-\frac{\partial }{\partial \chi }\left( \xi \eta
^{3}\right) \right] -ie^{-\alpha }\frac{\partial }{\partial \tau }\left(
e^{\alpha }\eta ^{2}\right) =0  \tag{A10}
\end{equation}%
\begin{equation*}
\frac{1}{\xi }\left[ \frac{\partial }{\partial \chi }\left( \xi \text{ }\eta
^{2}\right) -\frac{\partial \eta ^{1}}{\partial \theta }\right] -ie^{-\alpha
}\frac{\partial }{\partial \tau }\left( e^{\alpha }\eta ^{3}\right) =0.
\end{equation*}%
\noindent We define the following new components as
\begin{equation}
\eta ^{\pm }=\frac{1}{2}\left( \eta ^{1}\mp \eta ^{2}\right) =\left(
\begin{array}{c}
\eta _{+}^{jm}\left( \chi ;\tau \right) D_{+1,m}^{j}\left( \theta ,\varphi
\right)  \\
\eta _{-}^{jm}\left( \chi ;\tau \right) D_{-1,m}^{j}\left( \theta ,\varphi
\right)
\end{array}%
\right) ,  \tag{A11}
\end{equation}%
\noindent and

\begin{equation}
\eta ^{3}=\underset{jm}{4\pi \sum }\left( 2j+1\right) \eta _{0}^{jm}\left(
\chi ,\tau \right) D_{0,m}^{j}\left( \theta ,\varphi \right) .  \tag{A12}
\end{equation}%
\noindent \noindent Then the spinor $\eta $ becomes
\begin{equation*}
\eta \left( \chi ,\theta ,\varphi ;\tau \right) =\underset{jm}{4\pi \sum }%
\left( 2j+1\right)
\end{equation*}%
\begin{equation}
\times \left(
\begin{array}{c}
\lbrack \eta _{+}^{jm}\left( \chi ;\tau \right) D_{+1,m}^{j}\left( \theta
,\varphi \right) +\eta _{-}^{jm}\left( \chi ;\tau \right) D_{-1,m}^{j}\left(
\theta ,\varphi \right) ] \\
\eta _{0}^{jm}\left( \chi ,\tau \right) D_{0,m}^{j}\left( \theta ,\varphi
\right)  \\
\frac{1}{i}[\left( \eta _{+}^{jm}\left( \chi ;\tau \right)
D_{+1,m}^{j}\left( \theta ,\varphi \right) -\eta _{-}^{jm}\left( \chi ;\tau
\right) D_{-1,m}^{j}\left( \theta ,\varphi \right) \right) ]%
\end{array}%
\right) .  \tag{A13}
\end{equation}%
\noindent We substitute the expansion into equation and $\partial _{\pm }D$
\ properties of the $D$ functions: Then we obtain the following radial
equations:
\begin{equation*}
\frac{\sqrt{j\left( j+1\right) }}{\xi }\left( \eta _{+}^{jm}+\eta
_{-}^{jm}\right) =\frac{\partial }{\partial \tau }\left[ e^{-\alpha }\left(
e^{\alpha }\eta _{0}^{jm}\right) \right] ,
\end{equation*}%
\begin{equation*}
\frac{1}{\xi ^{2}}\frac{\partial }{\partial \chi }\left( \xi ^{2}\eta
_{0}^{jm}\right) =\frac{\sqrt{j\left( j+1\right) }}{\xi }\left( \eta
_{+}^{jm}-\eta _{-}^{jm}\right) ,
\end{equation*}%
\begin{equation}
\frac{1}{\xi }\frac{\partial }{\partial \chi }\left[ \xi \left( \eta
_{+}^{jm}+\eta _{-}^{jm}\right) \right] =e^{-\alpha }\frac{\partial }{%
\partial \tau }\left[ e^{\alpha }\left( \eta _{+}^{jm}-\eta _{-}^{jm}\right) %
\right] ,  \tag{A14}
\end{equation}%
\begin{equation*}
\frac{\sqrt{j\left( j+1\right) }}{\xi }\eta _{0}^{jm}-\frac{1}{\xi }\frac{%
\partial }{\partial \chi }\left[ \xi \left( \eta _{+}^{jm}-\eta
_{-}^{jm}\right) \right] =e^{-\alpha }\frac{\partial }{\partial \tau }\left[
e^{\alpha }\left( \eta _{+}^{jm}+\eta _{-}^{jm}\right) \right] .
\end{equation*}%
\noindent The combination of these equations give the following wave
equation in $1+1$ dimensions:
\begin{equation}
\left( \frac{\partial ^{2}}{\partial \chi ^{2}}-\frac{j\left( j+1\right) }{%
\xi ^{2}}-\frac{\partial ^{2}}{\partial \tau ^{2}}\right) \left( e^{\alpha
}\xi ^{2}\eta _{0}\right) =0.  \tag{A15}
\end{equation}%
This is the same with Eqs.(39) and the solutions of it the same with
Eqs.(40) and Eqs.(41).

\end{document}